\documentclass[aps,twocolumn,showpacs,superscriptaddress]{revtex4-2}

\usepackage{graphicx}
\usepackage{amsfonts}
\usepackage{amssymb}
\usepackage{amsbsy}
\usepackage{amsmath}
\usepackage{mathrsfs}
\usepackage{latexsym}
\usepackage{natbib}
\usepackage{bm}
\usepackage{subfigure} 
\usepackage{color}
\usepackage{wasysym}
\usepackage{mathbbol}
\usepackage{bigints}
\usepackage[font=small,labelfont=bf]{caption}
\allowdisplaybreaks

\begin{document}

	\title{Lorentzian path integral in Kantowski-Sachs anisotropic cosmology}

	\author{Saumya Ghosh}
	\email{sgsgsaumya@gmail.com, sg14ip041@iiserkol.ac.in}
	\author{Arnab Acharya}
	\affiliation{Department of Physical Sciences\\
		Indian Institute of Science Education and Research Kolkata, Mohanpur - 741246, WB, India}
	
	\author{Sunandan Gangopadhyay}
	\email{sunandan.gangopadhyay@gmail.com, \\sunandan.gangopadhyay@bose.res.in}
	\affiliation{Department of Astrophysics and High Energy Physics\\
		S.N. Bose National Centre for Basic Sciences, Salt Lake, Kolkata - 700106, India}
	
	\author{Prasanta K. Panigrahi}
	\affiliation{Department of Physical Sciences\\
		Indian Institute of Science Education and Research Kolkata, Mohanpur - 741246, WB, India}
	
	
	
	\begin{abstract}
		
		Motivated by the recent development in quantum cosmology, we revisit the anisotropic Kantowski-Sachs model in the light of a Lorentzian path integral formalism. Studies so far have considered the Euclidean method where the choice of the lapse integration contour is constrained by certain physical considerations rather than mathematical justification. In this paper, we have studied the Hartle-Hawking no-boundary proposal along with the use of Picard-Lefschetz theory in performing the lapse integration. In an isotropic limit, we show our results agree with the studies made in FLRW cosmology. We also observe that in the large scale structure the no-boundary proposal tends towards a conical singularity at the beginning of time. We have also performed a massless scalar perturbation analysis with no back reaction. This reveals that if there were any perturbation present at the beginning of the universe then that would flare up at the final boundary.

	\end{abstract}

	\maketitle
	
	\section{Introduction}\label{Introduction}
	
	One of the main motivations to study quantum cosmology is to understand the primordial universe and how it emerged from the Planck era to become the universe we observe today. Cosmology is governed by \emph{general relativity} (GR) but one needs to go beyond this classical theory to describe the small scale structure of the universe where a consistent quantum theory is supposed to have dominant effects. There are two main avenues to study quantum cosmology, the \emph{Wheeler-DeWitt} (WD) quantization approach \cite{DeWitt_PR_1967,Misner_PR_1969,Vilenkin_PRD_1994} and the path integral approach \cite{Teitelboim_PRD_1982,Hartle_PRD_1983,Halliwell_PRD_1989-I,Vilenkin_PLB_1982,Vilenkin_PRD_1988}. The WD quantization follows from the fact that the Hamiltonian operator annihilates the wavefunction of the universe. The Hamiltonian operator is achieved from the Hamiltonian constraint by replacing the canonical momenta of the field variables with their operator representation, $\pi^{ij} \rightarrow -i\hbar \frac{\delta~~}{\delta h_{ij} }$. However, there are issues associated with this approach. Firstly, there is an operator ordering ambiguity and secondly, there is a wide range of choices on the initial condition. On the other hand path integral method evaluates the probability amplitude of various final states of a system that has been prepared in a certain way as its initial state. So, in order to evaluate the transition probability of the universe one needs to have an idea about the initial state of the universe. 
	
	Hartle and Hawking's \emph{no-boundary} \cite{Hartle_PRD_1983,Hawking_NPB_1984,Halliwell_PRD_2019} proposal provides us with an initial condition of the universe. Their proposal dictates that the transition amplitude must be evaluated between a late time configuration of $3$-geometry and no initial $3$-geometry. This could help us understand how the universe was created from a zero volume $\sqrt{h}=0$ structure or `nothingness'.
	
	Anisotropic cosmological models are important in its own right. So far many studies have been made in the Wheeler-deWitt approach \cite{Sridip_PRD_2014,Pandey_AoP_2018,Saumya_MPLA_2019} as well as in the path integral quantization \cite{Dorronsoro_PRL_2018,Janssen_PRD_2019,Halliwell_PRD_1989}. The realization of no-boundary proposal gets fairly complicated as one needs to round off the big bang singularity with regularity conditions near the big bang for the Euclideanized spacetime. A general prescription has been presented in \cite{Halliwell_PRD_1989-II}. Kantwoski-Sachs model in particular is important because of its spatial topology, $S^{1} \times S^{2}$ which carries close resemblance with Euclidean black hole metrics  \cite{Sridip_CQG_2015,Saumya_MPLA_2022,Garcia-Compean_PRL_2002,Tucci_PRD_2019}.     
	
	According to the original proposal one needs to start with an Euclidean path integral to evaluate the transition amplitude. Also, the lapse integration is done along a complex contour constrained by certain physical considerations \cite{Halliwell_PRD_1990-II}. However, the recent proposal of \emph{Picard-Lefschetz} theory \cite{Feldbrugge_PRD_2017} gives a new way of choosing the lapse integration contour. The importance of this theory is that it allows one to carry out the Lorentzian path integral which has alluded theorists so far. It has been a long standing belief that the highly oscillatory nature of the Lorentzian path integral may not lead to a well defined result. 
	
	In this work our starting point is a Lorentzian path integral  with Kantowski-Sachs anisotropic cosmological model. Our study closely resembles the recent paper \cite{Fanaras_JCAP_2022}, where an Euclidean approach was followed to construct the wave function of the universe. However, there are few important distinguishing remarks presented in our analysis. It is well known that the half infinite range of the lapse function ($N=0$ to infinity) leads to the Green's function of the Wheeler-DeWitt equation. On the other hand a full infinite range yields the wave function. In this present work we focus on the half infinite range and proceed to evaluate the transition probability. Our choice of lapse contour on the complex $N$ plane is motivated by the Picard-Lefschetz theory. For a detailed discussion we refer \cite{Feldbrugge_PRD_2018}. In gravitational theory, working with imaginary time can lead to the \emph{conformal factor problem} - conformal transformation could make the Euclidean action arbitrarily negative \cite{Gibbons_NPB_1978}. So it is always prudent to work with the physically relevant Lorentzian time. We have also performed a perturbation analysis in the context of massless scalar field in the background of a classical spacetime. This reveals that the perturbations are unstable which reinstates the claims of \cite{Feldbrugge_PRL_2017}.
	
	The paper is organized as follows. In section \ref{Basic_formulation}, we briefly review the cosmological model and the gravitational theory along with some simplified calculations towards the path integral approach. Section \ref{Initial_conditions} contains the no-boundary initial conditions that we followed in our work. Section \ref{Contour_integration} deals with the detailed calculation and contour integration for the lapse function which leads to a final form of the path integral propagator. In section \ref{Scalar_field_fluctuation}, we have considered a massless scalar perturbation analysis with no back reaction. Finally we conclude in section \ref{Discussion}.

	\section{Basic formulation}\label{Basic_formulation}
	
	It is well known that the transition amplitude from an initial to a final state can be expressed as a path integral \cite{Feynman_McGraw-Hill_1965}. In the case of gravity the transition probability to propagate from an initial three spatial geometry $h_{ij}^{0}$ and matter field $\Phi_{0}$ to a final one $h_{ij}^{1}$, $\Phi_{1}$ is defined as
	\begin{equation}\label{Transition_amplitude}
		\mathcal{G} (h_{ij}^{1},\Phi_{1}|h_{ij}^{0},\Phi_{0}) = \int_{i}^{f} \mathcal{D}g_{\mu\nu}\mathcal{D}\Phi ~ e^{\frac{i}{\hbar}\mathcal{S}[g_{\mu\nu},\Phi]}~,
	\end{equation}    
	where $\mathcal{S}$ is the action for the metric $g_{\mu\nu}$ and the matter field $\Phi$. The path integral is taken over all possible values of four metrics $g_{\mu\nu}$ and matter field $\Phi$ with specified boundary conditions. The path integral defined above has Lorentzian signature. The standard trick to evaluate such an integral in gravity is to Euclideanize it. In this work, this will not be done and the approach in \cite{Feldbrugge_PRD_2017} will be followed.

	We begin by writing down the Einstein-Hilbert action in four dimensions with a positive cosmological constant 
	\begin{equation}\label{EH_action}
		\mathcal{S}[g_{\mu\nu}] = \frac{1}{2} \int_{\mathcal{M}} d^{4}\chi ~ \sqrt{-g}\big(\mathcal{R} - 2\Lambda\big) + \int_{\partial\mathcal{M}} d^{3}\eta ~ \sqrt{h}\mathcal{K}~,
	\end{equation}
	where $\mathcal{R}$ is the \emph{Ricci} scalar of the manifold $\mathcal{M}$, $\Lambda$ is the cosmological constant, $\sqrt{-g}$ is the determinant of the metric in coordinates $\{\chi\}$, $\{\eta\}$ are the coordinates on the three dimensional boundary $(\partial\mathcal{M})$ of the manifold, $\sqrt{h}$ is the determinant of the induced metric $h_{ab}$ on the boundary and $\mathcal{K}$ is the trace of the \emph{extrinsic curvature} tensor $\mathcal{K}_{ab}$ with respect to the induced metric. Here we have considered $8\pi G = 1$. The second term is called the \emph{Gibbon-Hawking-York} (GHY) term which ensures a well defined variational principle to yield the Einstein field equations of general relativity when the boundary geometries are held fixed.
	
	The invariant line element of Kantowski-Sachs (KS) metric with a spatial topology $S^{1}\times S^{2}$ follows
	\begin{equation}\label{KS_metric}
		ds^{2} = -\mathcal{N}^{2}(t)dt^{2} + a^{2}(t)dr^{2} + b^{2}(t)d\Omega_{2}^{2}~,
	\end{equation}
	where $\mathcal{N}$ is the \emph{lapse} function and $a,b$ are two scale factors in this anisotropic cosmological model, $r$ is a periodic coordinate with period $2\pi$ and $d\Omega_{2}^{2} = d\theta^{2} + \sin^{2}(\theta)d\phi^{2}$ is the metric on a unit two sphere with curvature $^{2}\mathcal{R} = 2$.   
	
	The action (\ref{EH_action}) along with the KS metric (\ref{KS_metric}) takes the form 
	\begin{equation}\label{Action_with_KS_metric}
		\mathcal{S}[a,b,\mathcal{N}] = \int_{t_0}^{t_1}~ dt \mathcal{L}(x,\dot{x}, \mathcal{N}) ~ +~ \mathcal{B}~,
	\end{equation}
	where $\mathcal{L}$ is the Lagrangian given by 
	\begin{equation}\label{Lagrangain_with_KS_metric}
		\mathcal{L} = \pi \Big( -\frac{2b\dot{a}\dot{b}}{\mathcal{N}} - \frac{a\dot{b}^{2}}{\mathcal{N}} - \mathcal{N}\Lambda ab^{2} + \mathcal{N}a \Big) ~.
	\end{equation}
	The only boundary ($\partial\mathcal{M}$) that we will consider is the final three surface at $t_{1}=1$. There is no boundary at $t_{0}=0$. If there was a boundary at both the ends then the GHY term in action (\ref{EH_action}) would cancel out and there will be no boundary term contribution. But in this case the boundary term at $t=0$ will be present and is given by
	\begin{equation}\label{Boundary_term_KS_metric}
		\mathcal{B}\big|_{t_{0}=0} = -\pi\Big(\frac{b^{2}\dot{a}}{\mathcal{N}} + \frac{2ab\dot{b}}{\mathcal{N}}\Big)_{t_{0}=0}~.
	\end{equation}

	\subsection{Lapse rescaling and variable change for simplified path integration}
	
	For mathematical simplification we now perform a rescaling in the lapse function as $\mathcal{N} = N/a$ and define $c = a^{2}b$. The action (\ref{Action_with_KS_metric}) then takes the form
	\begin{equation}\label{Action_KS_redfined_variables}
		\mathcal{S}[b,c,N] = \int_{0}^{1} dt \Big( - \frac{\dot{b}\dot{c}}{N} - N\Lambda b^{2} + N \Big)~.
	\end{equation}
	It can be seen that the Lagrangian (\ref{Action_KS_redfined_variables}) does not contain any $\dot{N}$ term which means $N$ is not a dynamical variable. Using the \emph{Batalin-Fradkin-Vilkovisky} quantization, one may impose the proper-time gauge $\dot{N} = 0$ \cite{Batalin_PLB_1977}. Without loss of generality one can choose the domain of $t$ within the range $[0,1]$. As derived by Halliwell \cite{Halliwell_PRD_1988} and Teitelboim \cite{Teitelboim_PRD_1982,Teitelboim_PRD_1983,Teitelboim_PRL_1983}, the propagator can be expressed as 
	\begin{eqnarray}\label{Transition_amplitude_KS_general}
		\mathcal{G}(b_{1}, c_{1} | b_{0}, c_{0}) = \int_{0}^{\infty} dN \int_{b}\mathcal{D}b\int_{c}\mathcal{D}c~ e^{\frac{i}{\hbar} (\mathcal{S}+\mathcal{B})} ~.
	\end{eqnarray} 
	Varying the action (\ref{Action_KS_redfined_variables}) with respect to $c$ and $b$ leads to the equations of motion 
	\begin{eqnarray}\label{EoM}
		\ddot{b} &=& 0 \nonumber \\
		\ddot{c} &=& 2N^{2}\Lambda b ~.
	\end{eqnarray}
	The \emph{Hamiltonian constraint} can be derived by varying the action with respect to $N$ which reads
	\begin{equation}\label{Hamiltonian_constraint}
		\frac{\dot{b}\dot{c}}{N} - N\Lambda b^{2} + N = 0 ~.
	\end{equation}
	The solution to the equations of motion (working with the gauge $\dot{N} = 0$) is given by
	\begin{eqnarray}\label{Solutions_of_EoM}
		\bar{b}(t) &=& (b_{1} - b_{0})t + b_{0} \nonumber \\
		\bar{c}(t) &=& \frac{N^{2}\Lambda (b_{1} - b_{0})}{3}t^{3} + N^{2}\Lambda b_{0}t^{2} \nonumber \\
		&+& \Big(- \frac{N^{2}\Lambda (b_{1} - b_{0})}{3} - N^{2}\Lambda b_{0} + c_{1} - c_{0}\Big)t + c_{0}~, \nonumber \\
	\end{eqnarray}
	where we have used the boundary values at $t_{0}=0$ and $t_{1}=1$ as 
	\begin{eqnarray}
		a(0) = a_{0}, ~~~ a(1) &=& a_{1} \nonumber \\
		c(0) = c_{0}, ~~~ c(1) &=& c_{1}~.
	\end{eqnarray}
	To perform a semiclassical approximation around the classical path, we now define 
	\begin{eqnarray}
		b(t) &=& \bar{b}(t) + X(t) \nonumber \\	
		c(t) &=& \bar{c}(t) + Y(t) ~.	
	\end{eqnarray}
	$\bar{b}(t)$ and $\bar{c}(t)$ are the saddle points of the action functional but they do not obey the Hamiltonian constraint. Putting this back in the action (\ref{Action_KS_redfined_variables}), we get 
	\begin{eqnarray}
		\mathcal{S}\big[X,Y,N\big] = \int_{0}^{1} dt \Big[&-&\frac{(\dot{\bar{b}}+\dot{X})(\dot{\bar{c}}+\dot{Y})}{N} \nonumber\\
		&-& N\Lambda(\bar{b}+X)^{2} + N\Big]~.
	\end{eqnarray}
	Using this form of the action in eq.(\ref{Transition_amplitude_KS_general}), the transition amplitude takes the form
	\begin{eqnarray}\label{Transition_amplitude_KS}
		\mathcal{G}(b_{1}, c_{1} | b_{0}, c_{0}) = \int_{0}^{\infty} dN e^{\frac{i}{\hbar} \big(\mathcal{S}_{0} + \mathcal{B}\big)} \int \mathcal{D}X\mathcal{D}Y e^{\frac{i}{\hbar}\mathcal{S}_{2}} ~,
	\end{eqnarray}
	where 
	\begin{eqnarray}
		\mathcal{S}_{0} = \pi \int_{0}^{1} dt \Big( - \frac{\dot{\bar{b}}\dot{\bar{c}}}{N} - N\Lambda \bar{b}^{2} + N \Big) \nonumber \\
		\mathcal{S}_{2} = \pi \int_{0}^{1} dt \Big(-\frac{\dot{X}\dot{Y}}{N} - N\Lambda X^{2}\Big) ~.
	\end{eqnarray}
	The boundary conditions on the $X,Y$ functional integration are $X(0)=0=Y(0)$ and $X(1)=0=Y(1)$.
	The final form of the classical action $\mathcal{S}_{0}$ in terms of the original scale factors, that is,  $a,b$ comes out to be
	\begin{equation}\label{Classical_action_original_variables}
		\mathcal{S}_{0} = \pi\Big[\alpha N - \frac{\beta}{N}\Big]~,
	\end{equation}
	with
	\begin{eqnarray}
		\alpha &=& 1 - \frac{\Lambda}{3}\big(b_{1}^{2} + b_{0}b_{1} + b_{0}^{2}\big)    \nonumber\\
		\beta &=& \big(b_{1} - b_{0}\big)\big(a_{1}^{2}b_{1} - a_{0}^{2}b_{0}\big) ~.
	\end{eqnarray}
	One can also look into a phase space form of the path integral. One then evaluates the conjugate momenta of $b$ and $c$
	\begin{eqnarray}
		\Pi_{b} = \frac{\partial \mathcal{L}}{\partial \dot{b}} &=& -\frac{\dot{c}}{N} \nonumber\\
		\Pi_{c} = \frac{\partial \mathcal{L}}{\partial \dot{c}} &=& -\frac{\dot{b}}{N}~.
	\end{eqnarray}
	The action in terms of the variables and their conjugate momenta reads
	\begin{equation}
		\mathcal{S} [x, \Pi, N] = \int_{0}^{1} dt \Big(\Pi_{b}\dot{b} + \Pi_{c}\dot{c} - N\mathcal{H}\Big)~, 
	\end{equation}
	where 
	\begin{equation}
		\mathcal{H} = -\Pi_{b}\Pi_{c} + \Lambda b^{2} - 1~.
	\end{equation}
	The Hamiltonian is independent of $c$ and linear in $\Pi_{b}$. So the semiclassical path integral can be done \emph{exactly}.

	\subsection{A separate set of variables for fixing the initial condition}
	For technical convenience let us perform a variable change $(a,b) \rightarrow (A,B)$ with 
	\begin{equation}\label{New_variables}
		A = b^{2} ~~~\text{and}~~~ B = ab ~.
	\end{equation} 
	In terms of these new variables, the Lagrangian (\ref{Lagrangain_with_KS_metric}) along with the lapse rescaling $\mathcal{N}=N/a$ takes the form
	\begin{eqnarray}\label{Lagrangian_new_variables}
		\mathcal{L} &=& \pi \Bigg( -\frac{B}{N A}\dot{A}\dot{B} +\frac{B^{2}}{4N A^{2}}\dot{A}^{2} - N\Lambda A  + N\Bigg) \\
		&\equiv& \frac{1}{2 N}f_{\gamma\sigma}\dot{q}^{\gamma}\dot{q}^{\sigma} - NU(q) ~. \nonumber	
	\end{eqnarray}
	The metric for the minisuperspace can be identified as
	\begin{eqnarray}\label{Minisuperspace_metric}
		f_{\gamma\sigma} = \begin{bmatrix}
			\frac{\pi B^{2}}{8 A^{2}} & -\frac{\pi B}{4 A} \\
			-\frac{\pi B}{4 A} & 0 \\
		\end{bmatrix}~.
	\end{eqnarray}
	The canonical conjugate momenta of the original variables in metric (\ref{KS_metric}) reads
	\begin{equation}\label{Conjugate_momenta_a_b}
		P_{a} = - \frac{2 b\dot{b}}{\mathcal{N}} ~~~\text{and}~~~ P_{b} = - \frac{2 b\dot{a}}{\mathcal{N}} - \frac{2 a\dot{b}}{\mathcal{N}} ~.
	\end{equation}
	Since we shall focus on fixing the initial values of the original variables, $a$ or $b$ and their Euclidean time derivatives, it will not be appropriate to work with $P_{a}$ and $P_{b}$. Instead a convenient choice would be the conjugate momenta $\Pi_{A}, \Pi_{B}$ corresponding to the variables $A,B$. In terms of $a, b$, they read
	\begin{equation}\label{Conjugate_momenta_A_B}
		\Pi_{A} = - \pi \frac{\dot{a}}{\mathcal{N}}  ~~~\text{and}~~~ \Pi_{B} = -2\pi \frac{\dot{b}}{\mathcal{N}} ~.
	\end{equation}
	
	\section{Initial conditions for No-Boundary proposal}\label{Initial_conditions}
	The original Hartle-Hawking proposal is that the path integral must be done on geometries that are \emph{compact} and the fields should be regular on such geometries. In the case of a positive cosmological constant $\Lambda$, any regular Euclidean solution of the field equations	is necessarily compact \cite{Hartle_PRD_1983}. Here we shall turn our attention to the smooth closure of the \emph{Euclidean} geometry. We change the metric signature of (\ref{KS_metric}) to an Euclidean time by taking the lapse convention $\mathcal{N} = i\mathcal{N}_{E}$ (the sign is chosen in accordance with the usual Wick rotation). The metric is then given by 
	\begin{equation}
		ds_{E}^{2} = \mathcal{N}_{E}^{2}(t)dt^{2} + a^{2}(t)dr^{2} + b^{2}(t)d\Omega_{2}^{2}~.
	\end{equation} 
	The starting condition is spatial volume be zero ($\sqrt{h} = 0$) at the initial time ($t=0$). As has been discussed in \cite{Halliwell_PRD_1990}, there could be two sets of conditions that can give a vanishing spatial volume. $a(0) = 0$ which corresponds to the closing of $S^{1}$ and $b(0) = 0$ which corresponds to closing of $S^{2}$. The corresponding regularity conditions are 
	\begin{eqnarray}\label{}
		a(0) = 0, ~ \frac{1}{\mathcal{N}_{E}}\frac{da}{dt}(0) &=&\pm 1, ~ \frac{1}{\mathcal{N}_{E}}\frac{db}{dt}(0) = 0  \\
		b(0) = 0, ~ \frac{1}{\mathcal{N}_{E}}\frac{da}{dt}(0) &=& 0, ~~~ \frac{1}{\mathcal{N}_{E}}\frac{db}{dt}(0) = \pm1 ~. 
	\end{eqnarray}
	Here we shall take up the boundary condition $a(0)=0$ along with $\frac{1}{\mathcal{N}_{E}}\frac{da}{dt}(0) = +1$, which, in the $A,B$ parametrization,  implies 
	\begin{eqnarray}
		B(0)&\equiv& B' = ab|_{t=0} = 0 \nonumber\\
		\Pi_{A}(0)&\equiv& \Pi_{A}' =  - \pi \frac{\dot{a}}{i\mathcal{N}_{E}} = i\pi~.
	\end{eqnarray}
	It is worth noting that we discarded the condition $\frac{1}{\mathcal{N}_{E}}\frac{db}{dt}(0) = 0$ since it would over constraint the theory and also in quantum mechanics one can not specify a variable and its canonical conjugate momenta at the same instant. 
	
	\section{Contour for lapse integration}\label{Contour_integration}
	We now proceed to specify the appropiate contour for carrying out the lapse integration. From eq.(\ref{Conjugate_momenta_A_B}) we have 
	\begin{eqnarray}
		\Pi_{A} = - \pi \frac{\dot{a}}{\mathcal{N}} ~.	
	\end{eqnarray}
	After the lapse rescaling $\mathcal{N} = N/a$, it takes the form
	\begin{eqnarray}
		\Pi_{A} = - \pi \frac{a\dot{a}}{N} ~.	
	\end{eqnarray}
	As has been discussed in the previous section, we will be fixing the initial values of $a$ and $\dot{a}$ at $t=0$. In the $A, B$ parametrization, this is equivalent to fixing $\Pi_{A}$ and $B$. Also the final point values of the scale factors are $a_{1}$ and $b_{1}$ at $t=1$. We take the help of classical solutions to find the value  of $b_{0}$ in terms of the boundary data $\Pi_{A}', B', a_{1}$ and $b_{1}$. Now at $t=0$
	\begin{eqnarray}
		\Pi_{A}' = -\pi\frac{\bar{a}\dot{\bar{a}}}{N} \Big|_{t=0} ~.
	\end{eqnarray}
	Using solutions of the equations of motion from eq.(\ref{Solutions_of_EoM}), one can find the following result
	\begin{eqnarray}\label{General_b_0}
		\frac{2N\Pi_{A}'}{\pi}b_{0} = \frac{N^{2}\Lambda}{3} \big( b_{1}+2b_{0} \big) - a_{1}^{2}b_{1} + \frac{{B'}^{2}b_{1}}{b_{0}^{2}} ~.
	\end{eqnarray}
	This is a cubic equation of $b_{0}$ which has three roots. But the initial boundary condition $B' = 0$ renders the above equation to a simple linear equation and $b_{0}$ comes out to be
	\begin{eqnarray}\label{b_0}
		b_{0} = \frac{b_{1}}{2N}\frac{\Big(\frac{N^{2}\Lambda}{3} - a_{1}^{2}\Big)}{\Big( \frac{\Pi_{A}'}{\pi} - \frac{N\Lambda}{3}\Big)}~.
	\end{eqnarray}
	The boundary term in eq.(\ref{Boundary_term_KS_metric}), in terms of lapse rescaling, as well as in $A,B$ parametrization reads
	\begin{eqnarray}
		\mathcal{B}\big|_{t=0} &=& -\pi\Big(\frac{b^{2} a\dot{a}}{N} + \frac{2a^{2}b\dot{b}}{N}\Big)_{t=0} \nonumber\\
		&=& A'\Pi_{A}' +B'\Pi_{B}'~.
	\end{eqnarray}
	Once again with the initial condition $B'=0$, the boundary term becomes
	\begin{eqnarray}\label{Boundary_term}
		\mathcal{B} = A'\Pi_{A}' = b_{0}^{2}\Pi_{A}' ~.
	\end{eqnarray}
	One can now use the value of $b_{0}$ in eq.(\ref{b_0}) and substitute it in eq.(\ref{Classical_action_original_variables}) along with eq. (\ref{Boundary_term}), which leads to
	\begin{eqnarray}\label{Final_form_of_action}
		\frac{\mathcal{S}_{0} + \mathcal{B}}{\pi} = N - \frac{\Lambda b_{1}^{2}}{3}N - \frac{a_{1}^{2}b_{1}^{2}}{N} - \frac{b_{1}^{2}}{4N^{2}}\frac{\big(\frac{\Lambda N^{2}}{3}-a_{1}^{2}\big)^{2}}{\frac{\Pi_{A}'}{\pi}-\frac{N\Lambda}{3}}~.~~~
	\end{eqnarray}
	The final form of the transition amplitude in terms of lapse integration is given by 
	\begin{equation}\label{Final_form_of_transition_probability}
		\mathcal{G} = \int_{0}^{\infty} dN ~ \mu\big(N\big) ~ e^{\frac{i}{\hbar}\mathcal{S}_{0} + \mathcal{B}} ~, 
	\end{equation}
	where the pre-factor $\mu\big( N \big)$ ensures normalization condition and is given by \cite{Schulman_Dover_2012}
	\begin{eqnarray}
		\mu(N) = f^{-\frac{1}{4}}\sqrt{|D|}{f'}^{-\frac{1}{4}}~,
	\end{eqnarray}
	where $f$ and $f'$ are the determinant of the minisuperspace metric $f_{\gamma\sigma}$ (\ref{Minisuperspace_metric}) evaluated at $t=1$ and $t=0$ respectively. Also, $D$ is the \emph{Van-Vleck-Morette} determinant and is given by \cite{Vleck_PNAS_1928,Morette_PR_1951}
	\begin{eqnarray}
		D = det\Bigg[\frac{\partial^{2}\tilde{S}}{\partial Q^{\gamma}\partial Z^{\sigma}}\Bigg]~,
	\end{eqnarray} 
	where $Q^{\gamma}=\{A,B\}$ and $Z^{\sigma}=\{\Pi_{A}',B'\}$. This leads to the following form of the prefactor
	\begin{eqnarray}\label{Prefactor}
		\mu\big( N \big) =\frac{4}{\pi} \Bigg[ \frac{2\pi A'}{N^{2}\big( \frac{\Pi_{A}'}{\pi} - \frac{N\Lambda}{3}\big)}\Bigg]^{1/2} ~.
	\end{eqnarray}
	One can perform a rescaling as follows
	\begin{equation}
		\Lambda N = \tilde{N} ,~ \Lambda a_{1}^{2} = u ,~ \Lambda b_{1}^{2} = v ~ \text{and} ~ \tilde{S} = \frac{\Lambda (\mathcal{S}_{0}+\mathcal{B})}{\pi} ~.
	\end{equation}
	Eq.(\ref{Final_form_of_action}) in terms of the rescaled quantities as mentioned above reads 
	\begin{equation}\label{Final_form_of_rescaled_action}
		\tilde{S} = \tilde{N} - \frac{\tilde{N}v}{3} - \frac{uv}{\tilde{N}} - \frac{v}{4\tilde{N}^{2}}\frac{(\frac{\tilde{N}^{2}}{3}-u)^{2}}{\frac{\Pi_{A}'}{\pi}-\frac{\tilde{N}}{3}} ~.
	\end{equation}
	Here we shall be working with the initial value $\Pi_{A}' = i\pi$. Also, we are interested in the large values of the scale factors. If we consider $v \gg 1$, then the above expression can be approximated to the following form
	\begin{equation}
		\tilde{S} = - \frac{\tilde{N}v}{3} - \frac{uv}{\tilde{N}} - \frac{v}{4\tilde{N}^{2}}\frac{(\frac{\tilde{N}^{2}}{3}-u)^{2}}{i-\frac{\tilde{N}}{3}} ~.
	\end{equation}
	In order to find an approximated value of the transition probability (\ref{Final_form_of_transition_probability}), we take help of the saddle point approximation. The saddle points can be found out by $\frac{d\tilde{S}}{d\tilde{N}} = 0$. There are five saddle points. Two of them are 
	\begin{eqnarray}
		\tilde{N}_{s} = \pm\sqrt{3u}~.
	\end{eqnarray}
	The other three are the roots of the equation
	\begin{eqnarray}
		-\tilde{N}_{s}^{3} + 6i\tilde{N}_{s}^2 + (12+3u)\tilde{N}_{s}-6iu = 0~.
	\end{eqnarray}
	These three roots are not relevant as we shall see  employing the \emph{Picard-Lefschetz} theory (Figure \ref{Steepest_descent_plot}). The only root that will contribute is $N_{s} = \sqrt{3u}$ as the steepest ascent contour passing through this point intersects the original domain of integration.
	
	This root has an interesting feature in our analysis as it makes the initial value of the scale factor $b_{0} = 0$, which is evident from eq.(\ref{b_0}). Here we already are working with the initial value $a_{0}=0$. However, zero initial condition for both the scale factors, $a,b$ leads to \emph{conical} singularity as mentioned in \cite{Halliwell_PRD_1990}. To circumvent this issue we consider a first order correction around this particular saddle point,   
	\begin{eqnarray}
		\tilde{N} = \tilde{N}_{s}(1 + \delta)~,	
	\end{eqnarray}
	where $|\delta|\ll 1$. With this correction, the exact action (\ref{Final_form_of_rescaled_action}) upto second order in $\delta$ takes the form 
	\begin{eqnarray}
		\tilde{S} = \sqrt{3u}\Big(1-\frac{2v}{3}\Big) + \delta + \frac{iv\delta^2}{\sqrt{3u}(\sqrt{3u}-3i)} + \mathcal{O}(\delta^{3})~.~~~
	\end{eqnarray}
	Extremizing the above action with respect to $\delta$ gives
	\begin{eqnarray}
		\delta = \frac{i(\sqrt{3u}-3i)}{2v}~.
	\end{eqnarray}
	It is worth noting that $\delta$ gets smaller as $v$ becomes larger. So for large values of $b_{1}$, this saddle point will tend to the value $\sqrt{3u}$.
	
	The action in terms of the saddle point value $\tilde{N} = \tilde{N}_{s}(1 + \delta)$ now reads
	\begin{eqnarray}
		\tilde{S} = i\frac{3u}{4v} + \frac{\sqrt{3u}(9+12v-8v^{2})}{12v}~.
	\end{eqnarray}
	In the complex $\tilde{N}$ plane, the general form of the integration (\ref{Final_form_of_transition_probability}) reads
	\begin{eqnarray}
		\mathcal{G} = \int_{\mathcal{C}} \mu(\tilde{N})e^{\lambda\tilde{S}(\tilde{N})} d\tilde{N} ~,
	\end{eqnarray}
	where $\lambda = i\pi/\hbar\Lambda$. For large values of $\lambda$, the asymptotic integral value can be given as \cite{k2005theory}
	\begin{eqnarray}
		\mathcal{G} \approx e^{\lambda \tilde{S}(\tilde{N}_{s})}\Bigg[\sqrt{\frac{2\pi}{\lambda |\tilde{S}''(\tilde{N}_{s})|}}~\mu(\tilde{N}_{s})e^{i\varphi_{m}} + \mathcal{O}(\lambda^{-3/2})\Bigg]~,\nonumber\\
	\end{eqnarray}
	where $\varphi_{m} = \frac{\pi-\theta_{0}}{2} + m\pi ~~(m=0,1)$ and $\theta_{0}=\text{arg}~\tilde{S}''(\tilde{N}_{s})$. The choice of value of $\varphi_{m}$ determines the sign in the formula, and naturally depends on the direction of integration along the contour $\mathcal{C}$. In our case $\lambda \propto 1/\hbar$ and we are interested in the semiclassical limit $\hbar \rightarrow 0$. The explicit form of the transition amplitude comes out to be
	\begin{eqnarray}
		\mathcal{G} \propto \frac{u^{\frac{1}{4}}}{v}\exp\Bigg[-\frac{3\pi u}{4\Lambda \hbar v} + i\Big(\frac{\pi}{\Lambda\hbar}\Big)\frac{\sqrt{3u}(9+12v-8v^{2})}{12v}\Bigg] ~.\nonumber\\
	\end{eqnarray}
	This can easily be recast in terms of the original variables by substituting $u=\Lambda a_{1}^{2}$ and $v=\Lambda b_{1}^{2}$. One can check that in the isotropic limit, that is, $a_{1}\approx b_{1}$, the real part in the classical action with the dominant saddle point contribution is negative and the exponential factor, $e^{-\frac{3\pi}{4\hbar\Lambda}}$  is very similar to the characteristic factor $e^{-\frac{12\pi^{2}}{\hbar\Lambda}}$ as derived in \cite{Feldbrugge_PRD_2017}. 
	\begin{widetext}
		\centering
		\includegraphics[height=10cm,width=13cm]{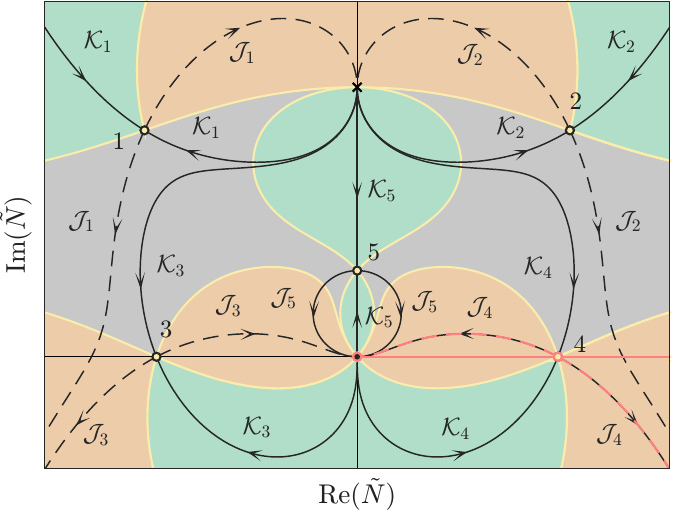}
		\captionof{figure}{The depiction of wedges and flow lines in the complex $\tilde{N}$ plane. There are total $5$ saddle points denoted by the numbers. Each saddle point has its steepest descent (Lefschetz thimbles $\mathcal{J}_{i}$) as well as steepest ascent ($\mathcal{K}_{i}$) flow lines passing through them. Steepest descent paths are marked with dashed lines and the solid lines indicate steepest ascent flows. The green colored region is where $Re[i\tilde{S}] > 0$ and the orange colored region denotes $Re[i\tilde{S}] < 0$. The grey colored region, however, is a degenerate region. Both steepest descent and ascent lines could pass through this region. The solid orange line along the positive real axis represents the original integration domain. The dashed orange line along $\mathcal{J}_{4}$ is the deformed contour. The fourth saddle point is the only contributing saddle as its steepest ascent contour $\mathcal{K}_{4}$ intersects the original integration domain. }
		\label{Steepest_descent_plot}
	\end{widetext}

	\section{Scalar field fluctuation}\label{Scalar_field_fluctuation}
	In this section we shall study the behaviour of an inhomogeneous, massless scalar field fluctuation in the anisotropic Kantowski-Sachs background. The action of the scalar field is given by
	\begin{eqnarray}\label{Scalar_action}
		S_{\Phi} = -\frac{1}{2}\int_{\mathcal{M}} d^{4}x \sqrt{-g} \nabla_{\!\!\mu}\Phi \nabla^{\mu}\Phi ~.
	\end{eqnarray}
	Using the Kantowski-Sachs metric with lapse rescaling, that is,
	\begin{eqnarray}
		ds^{2} = -\frac{N^{2}(t)}{a^{2}(t)}dt^{2} + a^{2}(t)dr^{2} + b^{2}(t)d\Omega_{2}^{2} ~,
	\end{eqnarray}
	one can rewrite the action (\ref{Scalar_action}) in the following form
	\begin{eqnarray}\label{Scalar_action_KS}
		S_{\Phi} = \frac{1}{2} \int_{0}^{1} dt N_{s}b^{2} \int_{0}^{2\pi}dr \int_{S^{2}}d\Omega_{2}\sqrt{g_{\Omega_{2}}}\Big[ \frac{a^{2}}{N_{s}^{2}}\big(\partial_{t}\Phi\big)^{2} \nonumber\\
		+ \frac{1}{b^{2}}\Phi\nabla^{2}\Phi \Big]~,~
	\end{eqnarray} 
	where the Laplacian is defined with respect to the metric, 
	\begin{eqnarray}
		d\Sigma^{2} = \frac{a^{2}(t)}{b^{2}(t)}dr^{2} + d\Omega_{2}^{2}~.
	\end{eqnarray}
	We now separate the scalar field into mode functions as follows 
	\begin{eqnarray}\label{Scalar_modes}
		\Phi(t,r,\theta,\phi) = \frac{1}{\sqrt{2\pi}}\sum_{\kappa,l,m}\varphi(t)e^{i\kappa r}Y_{lm}(\theta,\phi)~,
	\end{eqnarray}
	where the harmonics are labeled by the quantum numbers $\kappa, l$ and $m$. $Y_{lm}$ is related to the Legendre polynomial in the following way
	\begin{eqnarray}
		Y_{lm}(\theta,\phi) = \sqrt{\frac{(2l+1)(l-m)!}{4\pi(l+m)!}} (-1)^{m} e^{im\phi}P_{l}^{m}(\cos\theta)~.\nonumber 
	\end{eqnarray}
	The scalar field in terms of the mode functions (\ref{Scalar_modes}) follows the following eigenvalue equation
	\begin{eqnarray}\label{Eigen_eqn_Y_lm}
		\nabla^{2} \Phi = -\Big[l(l+1) + \frac{b^{2}}{a^{2}}\kappa\Big]\Phi~.
	\end{eqnarray}
	One thing to notice is that the eigenvalue is time dependent in this scenario. The orthogonality condition for $Y_{lm}$ reads 
	\begin{eqnarray}\label{Orthogonality_of_Y_lm}
		\int_{S^{2}} d\Omega_{2} Y^{\star}_{l'm'}(\theta,\phi)Y_{lm}(\theta,\phi) = \delta_{mm'}\delta_{ll'} ~.
	\end{eqnarray}
	It is worth mentioning that the scalar modes in (\ref{Scalar_modes}) forms a complete basis of complex functions. One can derive a complete basis set of real functions which along with the orthogonality condition (\ref{Orthogonality_of_Y_lm}) and the eigenvalue eq.(\ref{Eigen_eqn_Y_lm}) leads to the following form of the action (\ref{Scalar_action_KS})
	\begin{eqnarray}\label{Scalar_action_modes}
		S_{\Phi} = \sum_{\kappa,l,m} \int_{0}^{1} dt N_{s} \Bigg[\frac{a^{2}b^{2}}{2N_{s}^{2}} \big(\partial_{t}\varphi\big)^{2} - \frac{1}{2}\big[l(l+1) \nonumber\\
		+ \frac{b^{2}}{a^{2}}\kappa^{2}\big]\varphi^{2} \Bigg]~.
	\end{eqnarray}
	The equation of motion for each decoupled mode is given by
	\begin{eqnarray}\label{Scalar_EoM}
		\frac{d}{dt}\big[a^{2}b^{2}\dot{\varphi}\big] + N_{s}^{2}\big[l(l+1) + \frac{b^{2}}{a^{2}}\kappa^{2}\big]\varphi = 0~.
	\end{eqnarray}
	It is not possible to find an analytic solution for this equation. But eq.(\ref{Scalar_action_modes}) can be made simpler with the use of the equation of motion, and the on-shell action reads
	\begin{eqnarray}\label{On_shell_scalar_action}
		S_{\Phi}^{\text{on-shell}} = \frac{1}{2N_{s}}\sum_{\kappa,l,m}\Big[a^{2}b^{2}\varphi\dot{\varphi}\Big]_{t=0}^{1}~.
	\end{eqnarray}
	Here we proceed to solve the equation of motion (\ref{Scalar_EoM}) by numerical methods. We solve it for particular values of $l$ and $\kappa$ subjected to the boundary condition $\varphi(0) = 0$. We wish to keep $\varphi(1)=\varphi_{1}$ as a variable quantity. To do so, we start with a random initial value for $\dot{\varphi}(0)$ along with $\varphi(0)=0$ and find out $\varphi_{1}$. Then we divide the whole function $\varphi(t)$ with $\varphi_{1}$. Let us say the new function $\tilde{\varphi}(t) = \frac{\varphi(t)}{\varphi_{1}}$. So the boundary conditions on $\tilde{\varphi}(t)$ are $\tilde{\varphi}(0) = 0$ and $\tilde{\varphi}(1) = 1$. Now eq.(\ref{On_shell_scalar_action}) can be written down as
	\begin{eqnarray}
		S_{\Phi}^{\text{on-shell}} &=& \frac{1}{2N_{s}}a_{1}^{2}b_{1}^{2}\sum_{\kappa,l,m}\dot{\tilde{\varphi}}(1)\varphi_{1}^{2} \nonumber\\
		&\equiv& \sum_{\kappa,l,m} F(\alpha) \varphi_{1}^{2} ~,
	\end{eqnarray}  
	where we have defined $F(\alpha) = \frac{1}{2N_{s}}a_{1}^{2}b_{1}^{2}\dot{\tilde{\varphi}}(1)$ and $\alpha = a_{1}/b_{1}$. The transition probability, as we have derived for the background in the previous section, for the scalar field in a geometry $(\bar{a}(t),\bar{b}(t),N_{s})$ can be given as
	\begin{eqnarray}
		\mathcal{G}_{\Phi} \propto e^{\frac{i}{\hbar}S_{\Phi}^{\text{on-shell}}}~. 
	\end{eqnarray} 
	Below we analyze the nature of the imaginary part of $F(\alpha)$.

	\vspace{1cm}
	\includegraphics[height=8cm,width=8cm]{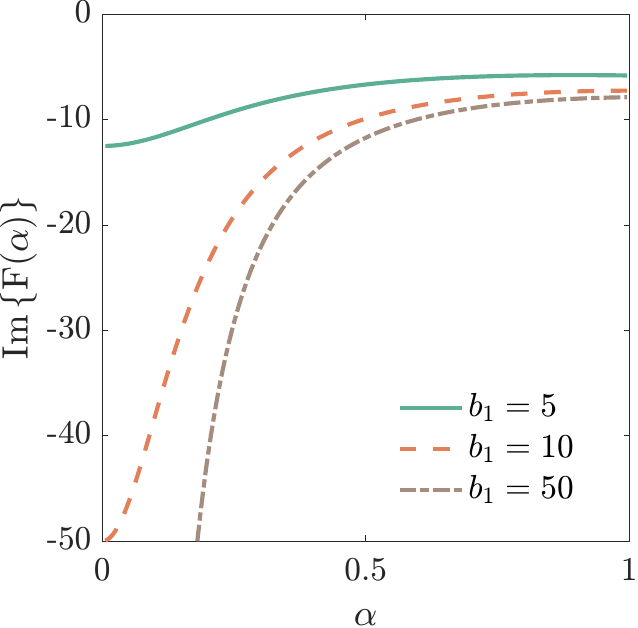}
	\captionof{figure}{In this picture we show the variation of imaginary part $F(\alpha)$ with $\alpha = \frac{a_{1}}{b_{1}}$ for a particular scalar mode \{$\kappa$,l,m\}. We see that the imaginary part of $F(\alpha)$ is always negative which implies that $\mathcal{G}_{\Phi}$ is always going to blow up as $\varphi_{1}$ increases. Near the isotropic limit, the imaginary part of $F(\alpha)$ reaches a constant value and in that region the real part of $\mathcal{G}_{\Phi}$ becomes an inverse Gaussian function which matches with the result obtained in case of FLRW cosmology \cite{Feldbrugge_PRL_2017}.}
	\vspace{1cm}
	\includegraphics[height=8cm,width=8cm]{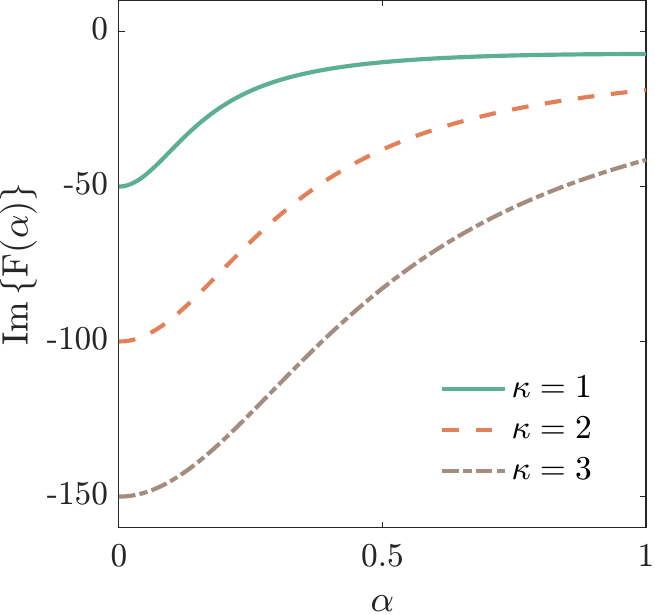}
	\captionof{figure}{The plot shows the variation imaginary part of $F(\alpha)$ with $\alpha$ for different $\kappa$ modes for a particular value of $l$. We see the larger values of $\kappa$ is going to blow up faster than the lower modes. }
	\vspace{1cm}
	\includegraphics[height=8cm,width=8cm]{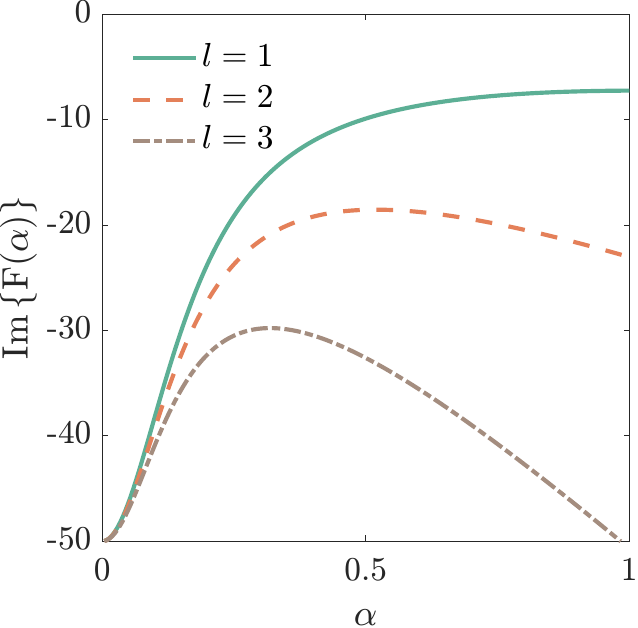}
	\captionof{figure}{The plot shows the variation of imaginary part of $F(\alpha)$ with $\alpha$ for different $l$ values while $\kappa$ remains fixed. Similar to the previous case, we see larger values $l$ makes the real part of $\mathcal{G}_{\Phi}$ blow up faster than the lower modes.}

	\section{Discussion}\label{Discussion}
	In this paper we considered a spatially homogeneous but anisotropic cosmological model whose spatial geometry is the product of a circle and a two sphere ($S^{1} \times S^{2}$). Our $4$D geometry has a preferred $3$D slicing which in turn ensures a convenient Euclidean time coordinate. Therefore, a simple rotation in the time flow leads the Euclidean ansatz to a Lorentzian one. In deriving a \emph{Hartle-Hawking} (HH) type solution one needs to specify the boundary data. The fixed boundary quantities in the path integral should be the same as that of in the classical variational principle. As has been delineated  in \cite{Halliwell_PRD_1990}, the appropriate boundary data for the variational principle is not the initial and final values of the scale factors $a,b$, but the final values of $a,b$ and initial value of $a$ along with Euclidean time derivative of $a$. Now the Euclidean propagation amplitude, $K_{E}\big(a_{1},b_{1}|a_{0},\dot{a}/N_{E}\big)$ can be considered analytic in its last argument (considering that all the mathematical intricacies have been resolved) and the Lorentzian counterpart can be defined as
	$K_{L}\big(a_{1},b_{1}|a_{0},i(\dot{a}/N)\big) \equiv K_{E}\big(a_{1},b_{1}|a_{0},\dot{a}/N_{E}\big)$. This is similar to the Wick rotation of Feynman propagator in quantum field theory. Our Lorentzian analysis is mainly motivated by this argument.   
	
	A general path integral prescription for anisotropic models was presented in \cite{Halliwell_PRD_1990}. Following that a recent work on tunneling formulation has been done in \cite{Fanaras_JCAP_2022} focusing on Kantowski-Sachs model in particular. These works are mainly based on an Euclidean path integral method where the choice of the lapse integration contour does not obey the Picard-Lefschetz theory. In this work we revisit the problem in the light of Picard-Lefschetz theory and go on to calculate the transition amplitude rather than the wave function of the universe. The importance of this work lies in the fact that we were able to carry out the Lorentzian path integral for the Kantowski-Sachs background by carefully tackling the saddle points arising in the Picard-Lefschetz theory. A neat result for the transition amplitude was obtained.
	
	We also performed a scalar perturbation analysis. This shows that if there were any initial perturbation then that would blow up as the universe grows bigger. This result matches with claims presented in \cite{Feldbrugge_PRL_2017} for FLRW cosmology. For large scale structure, we observed that the dominant saddle point contribution pushes the initial value of the scale factor, $b_{0}$ towards zero value. $b_{0} = 0$ is a singularity, that is, it violates the regularity condition at the beginning of time \cite{Halliwell_PRD_1990-II}. So it is safe to claim that the main contribution to the transition amplitude comes from an initial condition which tends to a singular start. This is in compliance with the perturbation analysis as both the results indicate the beginning of spacetime was not smooth. However, it will be interesting to elucidate this claim in further details from a general point of view.

	\begin{acknowledgments}
		S. Ghosh would like to thank Prof. Prasanta K. Panigrahi of IISER Kolkata for his support towards the completion of this project. 
	\end{acknowledgments}

\end{document}